\documentclass[10pt,conference,a4paper]{IEEEtran}
\usepackage{amsmath}
\usepackage{dblfloatfix}
\makeatletter
\newcommand{\eqnum}{\refstepcounter{equation}\textup{\tagform@{\theequation}}}
\makeatother
\usepackage{tabularx}
\makeatletter
\def\hlinewd#1{%
\noalign{\ifnum0=`}\fi\hrule \@height #1 %
\futurelet\reserved@a\@xhline}
\makeatother
\usepackage{cite} 
\usepackage[T1]{fontenc}
\usepackage[pdftex]{graphicx}
\usepackage[utf8]{inputenc}
\usepackage{url}

\usepackage{todonotes}

\usepackage{listings}
\begin{document}


\title{Implementation of BT-trees}

\author{\IEEEauthorblockN{Lars F. Bonnichsen} \IEEEauthorblockA{Technical University of Denmark\\
lfbo@dtu.dk}
\and
\IEEEauthorblockN{Christian W. Probst}
\IEEEauthorblockA{Technical University of Denmark\\
cwpr@dtu.dk}
\and
\IEEEauthorblockN{Sven Karlsson}\IEEEauthorblockA{Technical University of Denmark \\
svea@dtu.dk}
}


\maketitle

\begin{abstract}
This document presents the full implementation details of BT-trees, a highly efficient ordered map, and an evaluation which compares BT-trees with unordered maps.
BT-trees are often much faster than other ordered maps, and have comparable performance to unordered map implementations.
However, in benchmarks which favor unordered maps, BT-trees are not faster than the fastest unordered map implementations we know of.
\end{abstract}

\lstset{ %
escapechar=@,
  basicstyle=\ttfamily\footnotesize,        
  captionpos=b,                    
  frame=t,                    
  keepspaces=true,                 
  language=C++,                 
  showstringspaces=true,          
  tabsize=2,                       
}

\section{Introduction}
Unordered maps are a commonly used abstract data types, which store key-value associations: You can \emph{search} for which value is associated with a given key, \emph{insert} a new key-value pair association, or \emph{remove} a key-value pair association. 
\emph{0rdered maps} also support predecessor and successor searches, which return the next or previous key-value pair, according to an ordering over keys.

Ordered maps are typically implemented with balanced search trees or skiplists, while unordered maps are typically implemented with hash maps and tries.
Hash maps and tries are the most popular implementation choices, because they are fast in the general case. Unfortunately, hash maps and tries also have pathological cases, where they are slower than balanced search trees:
\begin{itemize}
\item Hash maps with poor hash functions have $O(n)$ running time; specifically operations on a hash map have an expected $O(n)$ running time if any bucket has $b$ key-value pairs associated, where $b \propto n$.
For instance, if 1 \% of all key-value pairs use the same bucket regardless of the number of key-value pairs, then the operations have an expected $O(n)$ running time. 
\item Tries are generally higher than balanced trees, especially when the key have many significant bits; for instance, a binary trie storing the $n$ first powers of two will have height $n$, whereas a balanced tree storing the same keys will have height $O(\log n)$.
\end{itemize}

To avoid the weaknesses of traditional ordered and unordered maps, we have developed BT-trees. BT-trees are ordered maps with similar performance to unordered maps, without pathological performance in corner cases. BT-trees are faster than traditional ordered maps, because they are designed for spatial locality, like B+trees, \emph{and} highly efficient operation implementations, unlike B+trees.
Compared to B+trees, BT-trees have a weaker balancing invariant and a simpler leaf node representation, which together permit a highly efficient implementation.




\section{BT-trees}

BT-trees are external, multiway, balanced search trees, \textit{ie} internal and leaf nodes have different representations, each internal node can have multiple children, and the has an $O(\log n)$ height.
BT-trees are intended to be used concurrently, by synchronizing with lock-elision: 
In the common case, \verb|acquire(lock)| and \verb|release(lock)| respectively correspond to starting a transaction, and committing a transaction while checking that the lock is not held.
If a transaction fails 2 times in a row, we instead acquire and release lock as a normal test-and-set lock.

Listing~\ref{lst:types} illustrates how BT-trees, key-value pairs, and
nodes are represented in pseudocode resembling C++.  The classes
\verb|I| and \verb|L| represent internal and leaf nodes respectively,
while \verb|E| represents key-value pairs.  Internal nodes have other
internal nodes or leaves as children. Leaf and internal nodes are
aligned to cache line boundaries by using the C++11 \verb|alignas|
keyword, and allocating with \verb|new|.  Each leaf node can store up
to $L_C$ key-value pairs, and internal nodes have up to $I_C$
children, where $L_C, I_C \ge 6$.  The lower bound node capacities of
$6$ ensure that we can split a full node into two nodes with at least
3 children or key-value pairs.  

\begin{figure}
\begin{lstlisting}[caption=Type definitions for BT-trees., label=lst:types]
class E<K, V> {K key; V value; }; // Key-value pairs

class alignas(64) L<K, V> { // Leaf nodes
  E<K, V> e[L_C]; // Unordered key-value pairs
}; 

class alignas(64) I<K> { // Internal nodes
  I* child[I_C]; // Pointers to children
  int size; // Number of children
  K key[I_C - 1]; // Internal node keys
};

class BT { // BT trees
  int height; // The tree's height
  I* root; // Pointer to the tree's root
  Lock lock; // The tree's lock
};
\end{lstlisting}
\vspace*{-5mm}
\end{figure}

\begin{figure}
\begin{lstlisting}[caption=Remove operation. Insert and search operations have the same structure, label=lst:rOps]
bool remove(const K& k, V& res, BT* t) {
  I* p, **pp; // Parent of leaf node
  sf16 ci; I* c; // leaf node
  while (1) { // 1. Find the leaf node
    findNode(k, pp, p, ci, c, t);
    // 2. Operate on the leaf node
    switch(remL(k, (L*) c, res)) {
    case SUCCESS:
      release(lock); return true;
    case FAILURE:
      release(t->lock); return false;
    case MERGE: // 3. Merge if near empty
      mergeL(pp, p, ci, c, t);
      release(t->lock); break;
    case SPLIT: // 3. Split if near full
      splitL(pp, p, ci, c, t);
      release(t->lock); break;
    }
  }
}
\end{lstlisting}
\vspace*{-7mm}
\end{figure}
~\\
~\\
All BT-tree operations follow the same template, as illustrated by Listing~\ref{lst:rOps}:
\begin{enumerate}
\item Find the leaf node which may hold the key.
\item Perform the operation on the leaf node.
\item If the leaf node is full or almost empty, split or merge it and try again.
\end{enumerate}


Listing~\ref{lst:lOps} illustrates how we search, insert, and remove from leaf nodes.
BT-trees split full leaf nodes when inserting into them, and merge non-root leaf nodes if they only have 3 key-value pairs when removing from them.
The operations iterate over up to $L_C$ key-value pairs in very simple loops, which can easily be unrolled manually, or by a compiler. We found that unrolling the \verb~srchL~ and \verb~insL~ is very beneficial, while \verb~remL~ benefits more from specializing the loops, to avoid looking for matching key-value pairs, or tracking the size of the leaf node, after a match has been found, or the size is confirmed to be above 3.

\begin{figure}
\begin{lstlisting}[caption=Operations on leaf nodes, label=lst:lOps]
Res srchL(const K& k, L* l, V& res) {
  for(int i = 0; i < L_C; i++) {
    E<K, V> e = l.e[uf32(i)]; // Look at all keys
    if (e.k == k) { // If we have a match
      res = e.v; // Return the matching value
      return SUCCESS;
    }
  }
  return FAILURE; // No matches in the leaf node
}

Res insL(const K& k, const V& v, L* l) {
  bool unused = false;
  int j; // Look at all keys
  for(int i = 0; i < L_C; i++) {
    E<K, V> e = l->e[i];
    if (e.k == 0) {
      unused = true;
      j = i; // Remember unused key-value pairs
    }
    if (e.k == k) {
      l->e[i] = {k, v}; // Replace any match
      return SUCCESS;
    }
  }
  if (unused) {
    l->e[j] = {k, v}; // Otherwise, replace any
    return SUCCESS; // empty key-value pair
  }
  return SPLIT; // Otherwise split the leaf node
}

Res remL(const K& k,  V& res, L* l) {
  bool match = false;
  int m, n = 0; // Look at all keys
  for(int i = 0; i < L_C; i++) {
    E<K, V> e = l->e[i];
    if(e.k != 0) {
      n++; // Track unused keys
    }
    if(e.k == k) {
      m = i; // Remember matching key
      res = e.v; // and value
      match = true;
    }
  }
  if(n <= 2 && !isRoot(l))
    return MERGE; // Merge nearly empty nodes
  if(match) {
    l->e[m].k = 0; // Remove matching key
    return SUCCESS;
  }
  return FAILURE; // No matching key
}
\end{lstlisting}
\vspace*{-7mm}
\end{figure}

Listing~\ref{lst:find} illustrates how we find the leaf node which may
hold a key \verb|k|: Preallocate nodes, begin the critical section,
and iteratively traverse from the root to the child which may
hold~\verb|k| until we reach a leaf node.  The internal nodes are
traversed by performing a linear search over the keys in the internal
nodes.  Any node with fewer than 3 children is merged, and any full
node is split.  After balancing nodes, the operation ends the
transaction and restarts.  In order to balance nodes we keep track of
the current node~\verb|c|, its parent~\verb|p|, and the pointer to the
parent, \verb|pp|.

\begin{figure}
\begin{lstlisting}[caption=Finding the leaf node which may hold the key $k$., label=lst:find]
void findNode(K k, I**& pp, I*& p, int& ci,
    I*& c, BT* t) {
start: // Allocate nodes before transactions
  ensureCapacity(6); 
  pp = &(t->r);
  p = 0; // The root has no parent
  acquire(t->lock);
  c = r; // Start at the root
  int h = t->h;
  if (h == 0)
    return; // The root is a leaf node
  int size = c->size;
  if (size == I::C) { // Split the root
    splitRoot(pp, c, h, size, t);
    release(t->lock); goto start;
  }
  p = c; ci = 0; // Traverse to child
  while(p->k[ci] <= k && ++ci != size - 1) {}
  c = p->c[ci];
  while (--h > 0) {
    size = c->size;
    if(size == 2) { // Merge small nodes
      mergeI(pp, p, ci, c, );
      release(t->lock); goto start;
    }
    if(size == I_C) { // Split full nodes
      splitInternal(c, p, pp, ci, size, t);
      release(t->lock); goto start;
    }
    pp = &p->c[ci]; // Traverse to child
    p = c; ci = 0; 
    while(p->k[ci] <= k && ++ci != size - 1) {}
    c = p->c[ci];
  }
}
\end{lstlisting}
\vspace*{-7mm}
\end{figure}

Splitting and merging nodes is handled by the same balancing function
given different arguments.  We split one node to produce two nodes
($in=1, out=2$), and we merge two nodes to produce one or two nodes
($in=2, out=1\vee out=2$).  Merging produces one output node when the
two input nodes have a combined size less than or equal to $b =
\frac{2}{3} (C + 2)$, where $C$ is the capacity of the output node
type, that is $L_C$ or $I_C$.  We chose between merging to one or
merging to two nodes in this way, because it maximizes the number of
operations required to bring the new nodes out of balance.  It takes
at least $C-b$ operations to fill a merged node, and at least $b-2$
operations to reduce a merged node's size to $2$. Similarly it takes
at least $C-\frac{b}{2}$ operations to fill a split node, and at least
$\frac{b}{2}-2$ operations to reduce a split nodes capacity to $2$.


Listing~\ref{lst:rebI} illustrates how internal nodes are balanced.
The key-value pairs of balanced leaf nodes are produced the same way
child pointers are produced for balanced internal nodes, except the
key-value pairs only have to be partially sorted: Copy the first $s =
\lceil ( c1.size + c2.size) / out \rceil$ elements from the unbalanced
node(s) to the first balanced node, \verb|n1|, and copy the remaining
elements to the second balanced node, \verb|n2|, if $out = 2$.  The
keys of balanced internal nodes are produced differently, because they
have to respect the order of the previous unbalanced nodes the key,
\verb|k|, which the parent of the previous nodes used to guide tree
searches: Order \verb|k| and the keys from the unbalanced nodes, and
copy the nodes into the balanced nodes such that the first balanced
node receives the first $\lceil ( c1.size + c2.size) / out \rceil - 1$
keys, and the second balanced node receives the last $\lfloor (
c1.size + c2.size) / out \rfloor - 1$ keys.  The $\lceil ( c1.size +
c2.size) / out \rceil - 1$'th key in the order is then inserted into
the parent of the balanced nodes, if $out=2$.  The actual
implementation uses more complicated code than~Listing~\ref{lst:rebI}
to avoid copying the keys and child pointers into intermediate arrays,
but is functionally equivalent.

\begin{figure}
\begin{lstlisting}[caption={Balancing internal nodes.}, label={lst:rebI}]
void balanceI(I* c1, I* c2, I* p, I** pp,
    int i, int in, int out, BT* t) {
  I* n1,*n2, I* p1; // The new nodes
  int s = ceiling((c1->size + c2->size) / out);
  auto ccomb = conc(c1->c, c2->c);
  auto kcomb = in == 2 ? c1->k :
      conc(c1->k, p->k[i], c2->k); // Gather the keys
  // Fill the new nodes
  memcpy(n1->c, ccomb, s * sizeof(void*));
  memcpy(n2->c, &ccomb[s],
      (c1->size + c2.size - s) * sizeof(void*));
  memcpy(n1->k, kcomb, (s - 1) * sizeof(K));
  memcpy(n2->k, &kcomb[s],
      (c1->size + c2.size - s - 1) * sizeof(K));
  p1.c[i] = (L*) n1; // Insert the balanced nodes
  if(out == 2) {
    p1.c[i + 1] = (L*) n2;
    p1.k[i] = n2.e[0].k;
  }
  int pSize = 0;
  if(p != 0) {
    pSize = p->size;
    ... copy p@{\color{gray}\verb~'~}@s other children and keys
    if(isRoot(p) && pSize == 2) {
      t->h--; // Merging the root
    }
  } else {
    t->h++; // Splitting the root
  }
  p1->size = pSize + out - in;
  *pp = p1; // Replace the parent
  dealloc(c1, c2, p);
}
\end{lstlisting}
\vspace*{-7mm}
\end{figure}

\section{Evaluation}

\begin{figure*}[tb]
\includegraphics[width=0.98\textwidth]{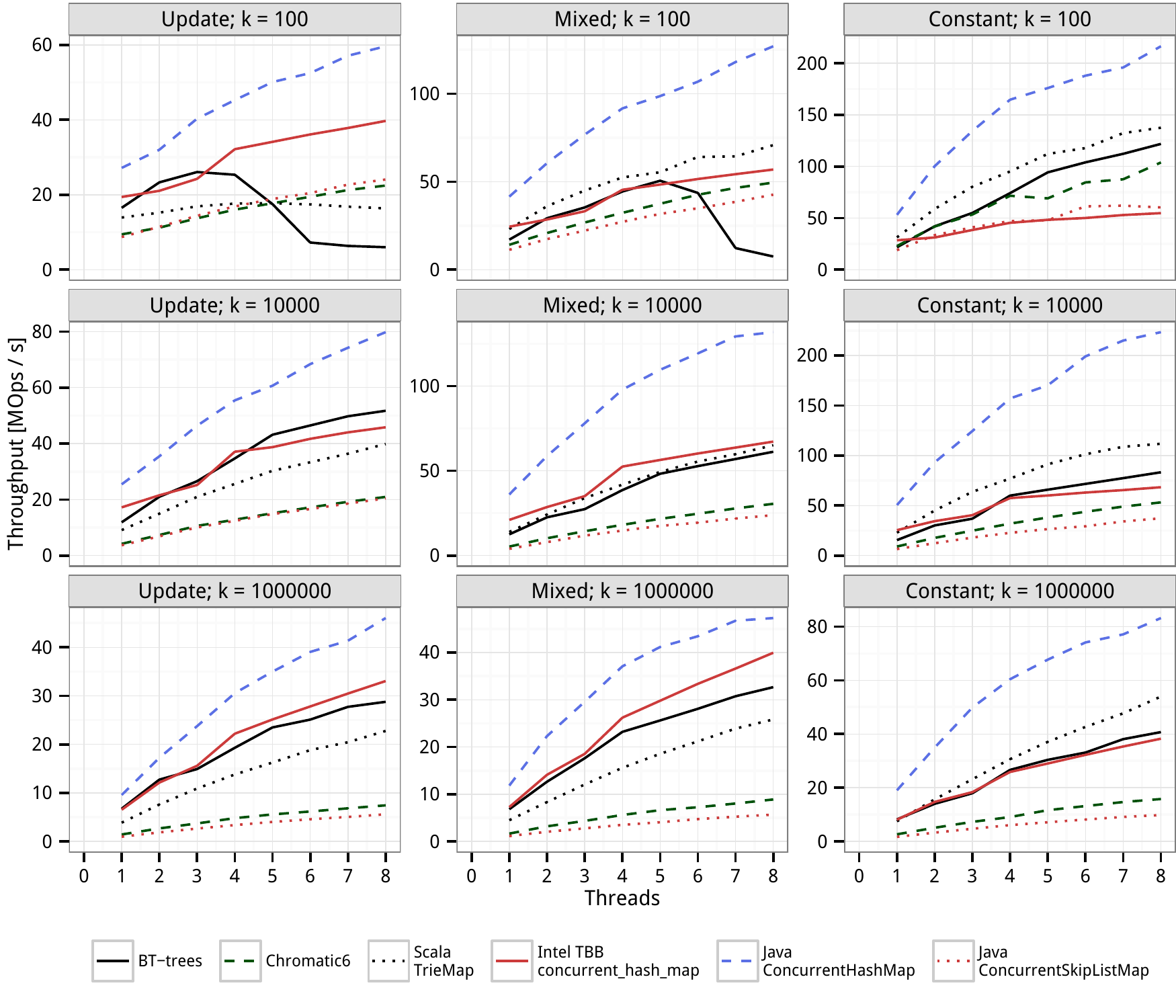}
\caption{Mean map throughput as a function of threads for 3 workloads and 3 key ranges.}
\label{fig:throughput2}
\end{figure*}

\begin{figure*}[tb]
\includegraphics[width=0.98\textwidth]{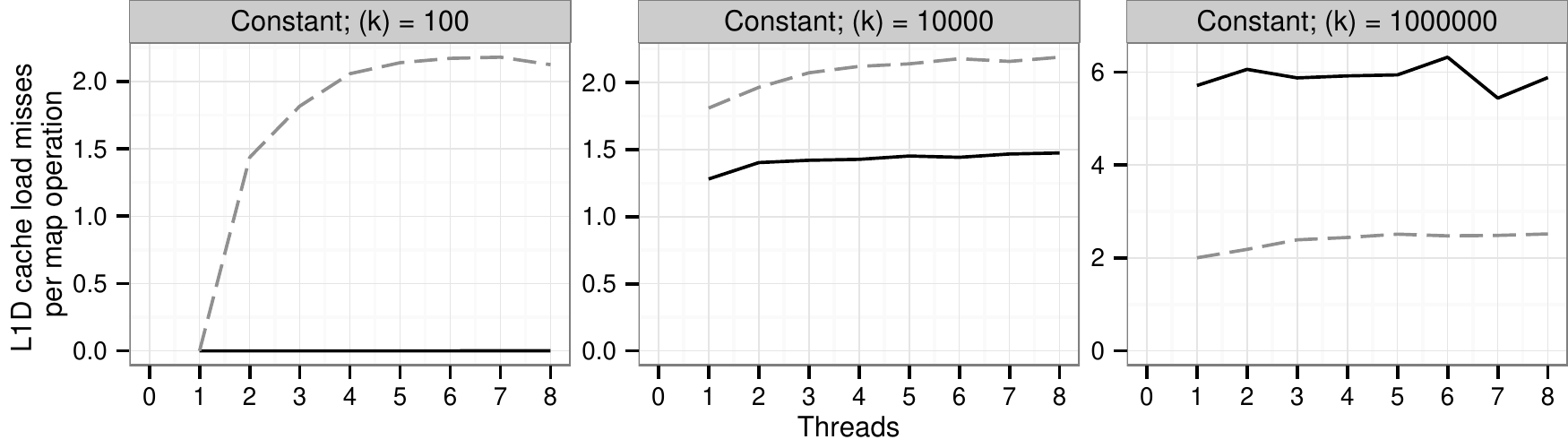}
\caption{The mean number of L1 cache load misses per operations for BT-trees (solid black line) and Intel TBB \texttt~concurrent\_hash\_map~ (dashed grey line). }
\label{fig:l1_load}
\end{figure*}

\begin{table}
\normalsize
\caption{Experimental machine}
\label{tab:machine}
\centering
\begin{tabular}{l|l}
Processor                   &  Intel Xeon E3-1276 v3@3.6GHz                \\
Processor specs                  & 4 cores, 8 threads \\
Processor specs(2) & 32KB L1D cache, 8 MB L3 cache  \\
C++ Compiler & GCC 4.9.1 \\
Java Compiler/Runtime & Oracle Server JRE 1.8.0\_20 \\
Operating system & Ubuntu Server 14.04.1 LTS \\
Kernel & 3.17.0-031700-generic \\
libc & eglibc 2.19
\end{tabular}
\end{table}

\begin{table}
\normalsize
\caption{Evaluated ordered (O) and unordered (U) maps}
\label{tab:maps}
\centering
\begin{tabular}{l|l}
Data structure name (Ordering) & Details \\\hline
BT-trees (O)                  & Node sizes $L_C=I_C=32$\\
Chromatic6 (O)~\cite{brown2014general} & Available online~\cite{brown}\\
ConcurrentSkipListMap (O)~\cite{javaSkipList} & Java (v1.8.0\_20)\\
ConcurrentHashMap (U)~\cite{javaCHM} & Java (v1.8.0\_20)\\
TrieMap (U)~\cite{Prokopec:2012} & Scala-library (v2.11.2)\\
concurrent\_hash\_map (U)~\cite{tbb} & Intel TBB (v4.3\_20141023) \\
\end{tabular}
\end{table}

This section covers our evaluation of BT-trees.
The evaluation compares the throughput of several ordered and unordered maps, on an established experiment, which avoids the weaknesses of unordered maps.
First we describe the experimental setup, then we discuss the implications of the experiments design, and finally we show and discuss the results of the experiments.

We evaluated BT-trees and the other maps listed in Table~\ref{tab:maps}, on the experimental machine described in Table~\ref{tab:machine}, with the map benchmark available at: \url{http://www.cs.toronto.edu/~tabrown/chromatic/testharness.zip}.
The benchmark is written in Java, so we had to port it to C++. 
In the benchmark, up to 8 threads repeatedly operate on one shared map for 5 seconds, after pre-filling the map with $n$ key-value pairs.
After the 5 seconds we recorded the number of operations performed on the map in the 5 seconds.
The map used 32 bit keys, and the operations used random keys, uniformly sampled from 1 to $k$, where $k$ is either 100,10,000, or 1,000,000.
We used 3 distributions of map operations:
\begin{enumerate}
\item \textit{Update}, with 50\% insertion, 50\% removal ($n = k / 2$);
\item \textit{Mixed}, with 70\% searches, 20\% insertion, and 10\% removal
  ($n = 2 k /3$); and
\item \textit{Constant}, with 100\% searches ($n = k$)
\end{enumerate}

The benchmark is designed to produce the highest possible throughput and contention for any given data structure size ($n$) and distribution of operations: The threads only generate keys and operate on the maps, unlike real applications which perform work do useful work between each map operation.
The maps are pre-filled with $n$ key-value pairs to minimize fluctuations in the maps size, and therefore minimize the changes in operation throughput during the benchmark; $n$ is
the expected number of key-value pairs in a map after infinitely many
operations.
The benchmark's design favors hash maps, because the keys have a very dense distribution.
A dense key distribution implies that most common integer hash functions are perfect hash functions.
In particular, the hash functions of the hash maps in Table~\ref{tab:maps} are perfect hash functions even when truncated to the least significant $log_2 (n)$ bits.
As a consequence, we expect the hash maps to have lower conflict rates, and higher throughput, than they would have for realistic inputs.
Despite being somewhat unrealistic, the benchmark has advantages: it is relatively well known, it is useful as a stress test, and it is a best case evaluation of hash maps.

Figure~\ref{fig:throughput2} shows the throughput of each map implementations on the benchmark.
The relative single threaded throughput of the implementations follow the same trend for all distributions of operations and keyranges:
\verb~ConcurrentHashMap~ is always faster than, BT-trees, \verb|TrieMap|, and
\verb~concurrent_hash_map~, which are usually faster than \verb|Chromatic6|, which in turn are usually faster than
\verb~ConcurrentSkipListMap~. 
There are 2 deviations from the usual
trend: (1) \verb|Chromatic6| are faster than
\verb~concurrent_hash_map~ in the \textit{Constant} workload when $k=100$, and
(2) \verb~ConcurrentSkipListMap~ achieves higher performance than
\verb|Chromatic6| in the \textit{Update} workload when $k=100$.  
The gap in performance between the traditional ordered maps, and the BT-trees / the unordered maps largest for large data structures (large $k$).
The increasing gap is caused by two factors: (1) BT-trees and \verb|TrieMap| being more cache efficient than traditional ordered maps, (2) hash maps have constant asymptotic running time, while skiplists have logarithmic asymptotic running time $O(1)$.
To illustrate the performance gap, BT-trees are 1.75, 2.84, and 4.71 times faster than \verb|Chromatic6| in the single threaded \textit{Update} workload for $k = 100$, 10,000, and 1,000,000, respectively.
In summary, BT-trees are slower than \verb~ConcurrentHashMap~, similar to \verb|TrieMap|, and
\verb~concurrent_hash_map~, and faster than the traditional ordered maps.

The relative performance of the map implementations is similar in parallel cases and the single threaded case.
Therefore we will focus on the gray area, the performance of BT-trees when compared to \verb|TrieMap| and \verb|concurrent_hash_map|.
\paragraph{BT-trees compared to \texttt{TrieMap}}
BT-trees are typically faster than \verb|TrieMap| in the \textit{Update} workloads, except when $k=100$, and slower in the \textit{Constant} workloads.
We believe that this is because the relative cost of insert / remove operations compared to search operations:
Insert and remove operations in BT-trees are performed in place on leaf nodes, and have similar costs to searching, while the \verb|TrieMap| insert and remove operations use copy-on-write, which increases their cost relative to search operations.

\paragraph{BT-trees compared to \texttt{concurrent\_hash\_map}}

BT-trees and \verb|concurrent_hash_map| have similar performance, except when $k=100$.
BT-trees scale poorly to multiple threads in the \textit{Mixed} and \textit{Update} workloads when $k=100$, but still achieves higher throughput than \verb~ConcurrentSkipListMap~ and \verb|Chromatic6|.
BT-trees poor scalability when $k=100$ is a side effect of using lock-elision; a side effect known as the Lemming effect~\cite{Dice:2009:techreport}.
Threads acquire the underlying lock when transactional executions of the critical section fails.
Acquiring the lock makes concurrent transactions more likely to fail:
Once a few transactions fail, many will transactions follow suit.



Meanwhile, \verb~concurrent_hash_map~ scales poorly in the \textit{Constant} workloads.
When $k=100$, \verb~concurrent_hash_map~ is the slowest
map in the \textit{Constant} workload. Figure~\ref{fig:l1_load} illustrates cache performance of BT-trees and \verb~concurrent_hash_map~ in the constant workloads. When going from 1 thread to 8 threads in \textit{Constant} $k=100$, \verb|concurrent_hash_map| execute more instructions per operation, and
cause up to up to 2.3 L1 cache load misses per operation.  By
comparison no other data structure we measured caused more than 0.01
L1 cache load misses per operation in the \textit{Constant} workload
with $k=100$.  The TBB \verb~concurrent_hash_map~ scales poorly in
this case because it uses a read-write lock per hash bucket.  search
operations acquire and release read locks by executing a
\verb|fetchAndAdd| atomic instruction.  The \verb|fetchAndAdd|
instructions, as well as any write instructions, invalidates the cache
lines of the other cores.  By comparison the other maps' search
operations do not write to the data structures memory.  The TBB
\verb~concurrent_hash_map~ is not significantly contended for larger
values of $k$, because then the hash map has more buckets, reducing
the risk of multiple threads searching adjacent buckets.

\section{Conclusion}
Traditional ordered maps are significantly slower than unordered maps, except in corner cases where the unordered maps have pathologically poor performance.
As an alternative, we present BT-trees, an ordered map, which has similar performance to unordered maps even in the best case scenario for unordered maps.
Specifically, BT-trees have similar performance to Intel TBB \verb|concurrent_hash_map| and Scala \verb|TrieMap|, but lower performance than Java \verb|ConcurrentHashMap|.

\bibliographystyle{ieeetr}

\bibliography{techReport}


\end{document}